\documentstyle[aps,12pt,manuscript]{revtex}
\begin{document}

\def\be{\begin{equation}}
\def\ee{\end{equation}}
\def\bea{\begin{eqnarray}}
\def\eea{\end{eqnarray}}

\def\pd{\partial}
\def\a{\alpha}
\def\b{\beta}
\def\g{\gamma}
\def\d{\delta}
\def\m{\mu}
\def\n{\nu}
\def\t{\tau}
\def\l{\lambda}

\preprint{KUNSAN-TP-01-2}

\title{Vibration modes of giant gravitons in the
background of dilatonic D-branes }

\author{ Jin Young Kim\footnote{Electronic address:
jykim@ks.kunsan.ac.kr}}
\address{Department of Physics, Kunsan National University,
Kunsan 573-701, Korea}
\author{ Y. S. Myung\footnote{Electronic address:
ysmyung@physics.inje.ac.kr}}
\address{Department of Physics, Inje University, Kimhae 621-749, Korea}
\maketitle

\begin{abstract}
We consider the perturbation of giant gravitons in the background
of dilatonic D-branes whose geometry is not of a conventional form
of ${\rm AdS}_m \times {\rm S}^n$. We use the quadratic
approximation to the brane action to investigate their vibrations
around the equilibrium configuration. We found the normal modes of
small vibrations of giant gravitons and these vibrations  are
turned out to be  stable.
\end{abstract}
\vfill

\newpage

Recently stable extended brane configurations in some string
theory background, called giant gravitons, attracted interests in
connection with the stringy exclusion principle. Myers
\cite{myers9910053} found that certain D-branes coupled to RR
potentials can expand into higher dimensional branes. McGreevy,
Susskind and Toumbas \cite{mst0003075} have shown that a massless
particle with angular momentum on the ${\rm S}$ part of ${\rm
AdS}_m \times {\rm S}^n$ spacetime blows up into a spherical brane
of dimensionality $n-2$. Its radius increases with increasing
angular momentum. The maximum radius of the blown-up brane is
equal to the radius of the sphere that contains it since the
angular momentum is bounded by the radius of ${\rm S}^n$. This is
a realization of the stringy exclusion principle \cite{sep}
through the AdS/CFT correspondence \cite{adscft}. Later it was
shown that the same mechanism can be applied to spherical branes
on the  AdS part \cite{gmt0008015,0008016}.
 However, they can grow arbitrarily large since there
is no upper bound on the angular momentum. To solve this puzzle,
instanton solutions describing the tunneling between the giant
gravitons on the AdS part and on the $S$ part
were introduced \cite{gmt0008015,jlee}. A magnetic analogue of
the Myers effect was investigated by Das, Trivedi and Vaidya
\cite{dtv0008203}. They suggested that the blowing up of gravitons
into branes are also possible on some backgrounds other than ${\rm
AdS}_m \times {\rm S}^n$ spacetime

Perturbations of the giant gravitons around their equilibrium
configuration were studied by Das, Jevicki and Mathur
\cite{djm0009019}. Using  quadratic approximation to the action,
they computed the natural frequency of the normal mode for giant
gravitons in ${\rm AdS}_m \times {\rm S}^n$ spacetime for both
cases when gravitons are extended in AdS space and they are
extended on the sphere. Frequencies are related to the curvature
scale of the background and are independent of the radius of the
brane itself. These modes are believed to play an important role
in the study of black holes within the string theory context. We
cannot explain the microscopic properties of a black hole if it is
regarded as a point singularity. But in string theory, a pointlike
singularity can be replaced by an extended brane, and vibration
modes of the brane can be used to study the microscopic entropy
\cite{micent} and Hawking radiation \cite{hawrad}.

In this paper, we will study the vibration mode of giant gravitons
in the near-horizon geometry of  dilatonic D-brane background. As
a concrete example, we will consider the case when a probe
$p$-brane wraps the transverse direction of ${\rm S}^{p+2}$ in the
near-horizon geometry of D$(6-p)$ branes \cite{dtv0008203}. Since
the background geometry is not exactly of the form ${\rm AdS}_m
\times {\rm S}^n$, it would be very interesting to study the
fluctuation analysis around this configuration.

Consider a probe D$p$-brane moving in the near-horizon geometry of
extremal $N$ D$(6-p)$-branes. The background metric is given by

 \be
 ds^2 = g_{tt} dt^2 + \sum_{k=1}^{6-p} g_{kk} (d x_k)^2 + g_{rr} dr^2
 + f(r) r^2 d \Omega_{p+2}^2 , \label{dsgen}
 \ee
where
 \be
 g_{tt} =  g_{kk} = \bigg ( {r \over R} \bigg )^{ {(p+1)} \over 2 } , ~~
 g_{rr} = f(r) = \bigg ( {R \over r} \bigg )^{ {(p+1)} \over 2 } ,
 \ee
 and $R$ can be expressed in terms of $N$ and the tension of a
 probe D$p$-brane as $ R^{p+1} = N/T_p V_p$, where
 $V_p = 2 \pi^{{p+1} \over 2} / \Gamma ( {{p+1} \over 2} )$ is the
 volume of unit ${\rm S}^p$.
$d \Omega_{p+2}^2$ can be parametrized as

 \be
 d \Omega_{p+2}^2 = {1 \over {1 - \rho^2}} d \rho^2 +  (1 - \rho^2) d \phi^2
    +  d \Omega_p^2 ,
 \ee
with

 \be
 d \Omega_{p}^2 = d \theta_1^2 + \sin^2 \theta_1 ( d \theta_2^2 +
 \sin^2 \theta_2 ( \cdots + \sin^2 \theta_{p-1} ) d \theta_p^2 ) .
 \ee
Also we have the  dilaton ($\Phi$) and RR potential ($A^{p+1}$) given by

 \bea
 e^{\Phi} &=& \bigg ( {R \over r} \bigg )^{ {(p-3)(p+1)} \over 4 } ,
  \label{dil}    \\
 A_{\phi \theta_1 \dots \theta_p}^{p+1} &=& R^{p+1} \rho^{p+1}
    \epsilon_{\theta_1 \dots \theta_p} , \label{RR}
 \eea
where $ \epsilon_{\theta_1 \dots \theta_p}$ is the volume form of
the unit $p$-sphere. In the case of $r=R$, $\Phi=0$. For $0<r<R$,
we have a non-zero dilaton. In this sense, Eq.(\ref{dsgen}) is
called the dilatonic D-brane background.

We consider  an equilibrium configuration in which a probe  D$p$-brane
wraps the ${\rm S}^p$. The action for this case,
ignoring the fermions, is given by

 \bea
 S &=& S_{DBI} + S_{CS} \nonumber  \\
   &=&-T_p \int d^{p+1}\xi\,e^{-\Phi}\sqrt{-{\rm
 det}\,\hat{G}_{\alpha\beta}}+ T_p\int d^{p+1}\xi\,\hat{A}^{p+1},
 \label{bract}
 \eea
where $\hat{G}_{\alpha\beta}$ and $\hat{A}^{p+1}$ are
 pullbacks of the metric and the RR $(p+1)$-form potential,
 respectively

 \bea
 {\hat G}_{\alpha\beta}&=&
 { {\partial x^M } \over {\partial \xi^\alpha} }
      {  {\partial x^M } \over {\partial \xi^\beta} } g_{MN}  ,  \\
 {\hat A}^{(p+1)}_{\xi_0 \xi_1 \dots \xi_p}
 &=& A^{(p+1)}_{M_1 \dots M_p}
  { {\partial x^{M_1}} \over {\partial \xi_0}}
  { {\partial x^{M_2}} \over {\partial \xi_1}}
  { {\partial x^{M_{p+1}} } \over {\partial \xi_{p}} } .
 \eea
If we choose a static gauge : the time parameter of the
worldvolume $\xi^0 = t$ ; the $p$ angular parameters $\xi_i$ are
set to be the angles on  $S^p$, $\xi_i = \theta_i$, then the
dynamical variables are given by $r(t,\theta_i), x_k (t,
\theta_i), \rho (t,\theta_i)$ and $\phi (t,\theta_i)$. We choose
an equilibrium configuration such that these quantities do not
depend on $\theta_i$ so that there are no brane oscillations.
Since there exist translational symmetries along $x^i$, the
corresponding momenta are conserved. We will study the motion
whose conserved momenta are identically zero. Then, the dynamical
variables take the form of  $r(t), \rho(t)$ and $\phi(t)$. With an
appropriate choice of gauge, we get the full probe brane action as

 \be
 S = - T_p V_{p} \int dt e^{-\Phi} (f(r) \rho^2 r^2)^{p/2}
\sqrt{g_{tt}-g_{rr}{\dot r}^2-g_{\rho\rho}{\dot \rho}^2 - g_{\phi
\phi}{\dot \phi}^2 }
 + N \int dt  \rho^{p+1} {\dot \phi},
 \ee
where dot denotes  derivative with respect to $t$ and $V_p$
stands for the volume of  unit $p$-sphere.
Its conjugate momenta and  Hamiltonian are:

 \bea
 P_r &=& {\partial L \over \partial {\dot r}}
 = {T_p V_p e^{-\Phi} \over \sqrt{g_{tt}-g_{rr}{\dot
 r}^2-g_{\rho\rho}{\dot \rho}^2 - g_{\phi \phi}{\dot \phi}^2 }  }
  (f \rho^2 r^2)^{p/2} g_{rr} {\dot r} , \\
 P_{\rho}&=& {\partial L  \over \partial {\dot \rho}}
 = {T_p V_p e^{-\Phi} \over \sqrt{g_{tt}-g_{rr}{\dot r}^2-g_{\rho\rho}{\dot
 \rho}^2 - g_{\phi \phi}{\dot \phi}^2}}
            (f \rho^2 r^2)^{p/2} g_{\rho\rho} {\dot \rho} , \\
 P_{\phi}&=&{\partial L  \over \partial {\dot \phi}}=
 {T_p V_p e^{-\Phi} \over \sqrt{g_{tt}-g_{rr}{\dot r}^2-g_{\rho\rho}{\dot
 \rho}^2 - g_{\phi \phi}{\dot \phi}^2} }
  (f \rho^2 r^2)^{p/2} g_{\phi \phi} {\dot \phi} + N \rho^{p+1} ,
  \label{momphi}
 \eea

 \bea
 H&=&P_r{\dot r} + P_{\rho}{\dot \rho} + P_{\phi}{\dot \phi} - L
    \nonumber \\
  &=&\sqrt{g_{tt}} \bigg [ (T_p V_p e^{-\Phi})^2 (f(r) \rho^2 r^2)^{p}
+ {P_r^2 \over g_{rr}} + {P_{\rho}^2 \over g_{\rho\rho}}+
{(P_{\phi}-N\rho^{p+1})^2 \over g_{\phi\phi}} \bigg ]^{1/2} .
  \label{hameq} \eea
Note that if we impose a condition
 \be
 T_p V_p e^{-\Phi} (f(r) r^2)^{{p+1 \over 2}} = N, \label{tvn}
 \ee
 and use $g_{\phi\phi}=f(r) r^2 (1-\rho^2)$, we can combine
  the first and last terms within the
square brackets of Eq. (\ref{hameq}). Then the Hamiltonian can be
expressed as

 \be
 H=\sqrt{g_{tt}} \bigg [ {P_{\phi}^2 \over f(r) r^2} + {P_r^2 \over
g_{rr}} +{P_{\rho}^2 \over g_{\rho\rho}}+ {(\rho
P_{\phi}-N\rho^p)^2 \over g_{\phi\phi}} \bigg ]^{1/2}.
 \ee
Since the Lagrangian does not depend on $\phi$, $P_{\phi}$ is a
constant of motion. For a given $P_{\phi}$, the lowest energy
configuration satisfies $P_{\rho} = 0$ for all time because $\rho$
does not appear in the first two terms. Here we find  a relation
between the equilibrium value of $\rho_0$ and  $P_{\phi}$

 \be
  P_{\phi} = N \rho_0^{p-1} , \label{pphieq}
 \ee
 which is obviously independent of the radial coordinate $r$.
The corresponding  Hamiltonian  is

 \be
 H=\sqrt{g_{tt}} \bigg [ {P_{\phi}^2 \over f(r) r^2} + {P_r^2 \over
 g_{rr}} \bigg ]^{1/2}.
 \ee
Actually this is the Hamiltonian of a massless particle with
angular momentum  $P_{\phi}$ on  ${\rm S}^{p+2}$ sphere. Unlike
the usual massless particle, the angular momentum of a probe brane
is bounded because of  $0 \le \rho \le 1$. A possible maximum
value of its angular momentum is $N$. This reminds us  the stringy
exclusion principle. Eq.(\ref{tvn}) is an important condition for
the brane to behave like a massless particle.

So far we reviewed how the configuration of giant graviton appears
in the near-horizon background of dilatonic D-brane. Now we
consider the small vibration of  giant graviton around the stable
equilibrium configuration. In general, the brane  can move in any
direction. We can consider the brane motion along any of the $r$,
$\rho$ and $\phi$. If one considers the case when the brane moves
along the $r$ direction, one can get the cosmological model known
as the mirage cosmology \cite{migcos}. The motion of the probe (or
universe) brane in ambient space induces cosmological expansion
(or contraction). And if we consider the brane motion along the
$\rho$ and $\phi$ directions, we get the giant graviton picture as
in Ref. \cite{dtv0008203}. Since we are interested in the
perturbation from stable configuration of the giant graviton, we
neglect the motion of the probe brane along the transverse
direction ($r$) of the background D$(6-p)$-branes. So we set $\dot
r = 0$, i.e. $r = r_0 ={\rm constant} $. We find the angular
velocity of this configuration from Eqs. (\ref{momphi}) and
(\ref{pphieq})

 \be
 \dot \phi \equiv \omega_0 = \pm {1 \over {f_0 r_0} }
 \ee
with $f_0 = f (r_0) = (r_0 /R)^{(p+1)/2}$.

A small vibration of the brane can be  described by defining
spacetime coordinates $(\rho,\phi,x_k)$ as a function of
the worldsheet coordinates
$\xi_0 , \xi_1 , \cdots \xi_p $. In the static gauge, where

 \be
 \xi^0 = t = \tau , ~~~ \xi_i = \theta_i , ~~~ i= 1, \cdots , p ,
 \ee
 perturbation of the remaining coordinates can be written as

 \bea
 \rho &=& \rho_0 + \epsilon~ \delta \rho (\tau, \xi_1, \dots ,
 \xi_p) ,  \\
 \phi &=& \omega_0 \tau + \epsilon~ \delta \phi (\tau, \xi_1, \dots ,
 \xi_p) ,  \\
 x_k &=&  \epsilon~ \delta x_k (\tau, \xi_1, \dots , \xi_p) ,~~~
 k = 1, \dots , 6-p .
 \eea

First we expand the action (\ref{bract}) to linear-order in
$\epsilon$

 \bea
 S(\epsilon) &=& -\epsilon~T_p \int d\tau d^p \xi \sqrt{g_\xi}
 e^{-\Phi_0} (f_0 r_0^2)^{p \over 2} \rho_0^{p-1}  \nonumber \\
  && ~~~~~~\times \bigg [
 { {(p+1) f_0 r_0^2 \omega_0^2 \rho_0^2
    + p (1/f - f_0 r_0^2 \omega_0^2 ) } \over
 \sqrt{1/f - f_0 r_0^2 (1 - \rho_0^2) \omega_0^2} } \delta \rho
  -  { {(1-\rho_0^2) f_0 r_0^2 \omega_0 \rho_0 } \over
 \sqrt{1/f - f_0 r_0^2 (1 - \rho_0^2) \omega_0^2} } \delta {\dot
 \phi}  \bigg ]  \nonumber  \\
&+& \epsilon N \int d\tau \rho_0^p
    [ (p+1) \omega_0 \delta \rho + \rho_0 \delta {\dot \phi} ].
 \eea
Using Eq.(\ref{tvn}), we have

 \bea
 S(\epsilon) = -  \epsilon N \rho_0^{p-1}  \int d\tau
  && \bigg [
  \bigg \{  {1 \over \sqrt{f_0 r_0^2} } { {(p+1) f_0 r_0^2 \omega_0^2 \rho_0^2
    + p (1/f - f_0 r_0^2 \omega_0^2 ) } \over
 \sqrt{1/f - f_0 r_0^2 (1 - \rho_0^2) \omega_0^2} } -
  (p+1) \rho_0 \omega_0  \bigg \} \delta \rho  \nonumber  \\
 && + \bigg \{ {1 \over \sqrt{f_0 r_0^2} }
     { {- (1-\rho_0^2) f_0 r_0^2 \omega_0 \rho_0 } \over
 \sqrt{1/f - f_0 r_0^2 (1 - \rho_0^2) \omega_0^2} } - \rho_0^2
 \bigg \} \delta {\dot
 \phi}  \bigg ] .
 \eea
Clearly the coefficient of $\delta \rho$ vanishes if one uses
 $\omega_0 = \pm {1 \over {f_0 r_0} }$. Also the coefficient
of $\delta {\dot \phi}$ is a constant and  thus  this term does
not contribute to the variation of the action upon the integration
over $\tau$.

On the other hand, the second order term in $\epsilon$ is calculated as

 \bea
 S(\epsilon^2) &=& \epsilon^2{ N \over T_p} \omega_0 \rho_0^{p-1}
  \int d\tau d^p \xi \sqrt{g_\xi}   \nonumber \\
 \times && \bigg[
    {1 \over 2} { 1 \over {\omega_0^2 (1 - \rho_0^2) } }
    ( \delta {\dot \rho} )^2
 -  {1 \over 2} { 1 \over { (1 - \rho_0^2) } }
 {{\partial \delta \rho} \over {\partial \xi_i} }
  {{\partial \delta \rho} \over {\partial \xi_j} }
 g^{\xi_i \xi_j}                    \nonumber \\
 &&+ {1 \over 2} { {1 - \rho_0^2} \over {\omega_0^2 \rho_0^2 } }
 ( \delta {\dot \rho} )^2
 -  {1 \over 2} (1 - \rho_0^2)
 {{\partial \delta \phi} \over {\partial \xi_i} }
  {{\partial \delta \phi} \over {\partial \xi_j} }
 g^{\xi_i \xi_j}                    \nonumber \\
 && + {{p-1} \over {\omega_0 \rho_0} }
      \delta \rho \delta {\dot \rho}   \nonumber \\
 && + {1 \over 2} \sum_{k=1}^{6-p}
      (\delta {\dot x_k} )^2
 -  {1 \over 2} { 1 \over { f_0 r_0^2 } }
  \sum_{k=1}^{6-p}
 {{\partial \delta x_k} \over {\partial \xi_i} }
  {{\partial \delta x_k} \over {\partial \xi_j} }
 g^{\xi_i \xi_j}   \bigg ] .
 \eea
The equations of motion are

 \bea
 && {1 \over {\omega_0^2 (1 -\rho_0^2) } }
 { {\partial^2 \delta \rho} \over {\partial \tau^2} }
 -  {1 \over {1 -\rho_0^2 } }
 {\partial \over {\partial \xi_i} } \bigg (
  {{\partial \delta \rho} \over {\partial \xi_j} }
 g^{\xi_i \xi_j}  \bigg )
 - {{p-1} \over {\omega_0 \rho_0 } }
 { {\partial \delta \phi} \over {\partial \tau} }
 = 0,  \label{eofmrho} \\
 && {{1- \rho_0^2} \over {\omega_0^2 \rho_0^2 } }
{ {\partial^2 \delta \phi} \over {\partial \tau^2} }
 -  (1 -\rho_0^2 )
 {\partial \over {\partial \xi_i} } \bigg (
  {{\partial \delta \phi} \over {\partial \xi_j} }
 g^{\xi_i \xi_j}  \bigg )
 + {{p-1} \over {\omega_0 \rho_0 } }
 { {\partial \delta \rho} \over {\partial \tau} }
 = 0,  \label{eofmphi}  \\
 && { {\partial^2 \delta x_k} \over {\partial \tau^2} }
 -  {1 \over {f_0^2 r_0^2 } }
 {\partial \over {\partial \xi_i} } \bigg (
  {{\partial \delta x_k} \over {\partial \xi_j} }
 g^{\xi_i \xi_j}  \bigg ) = 0 .  \label{eofmxk}
 \eea
We observe that $\delta x_k$ perturbations are decoupled from $\delta \rho$
and $\delta \phi$.
Let us Introduce the spherical harmonics $Y_l$ on ${\rm S}^p$,

 \be
 g^{\xi_i \xi_j}
  { \partial \over {\partial \xi_i} }
  { \partial \over {\partial \xi_j} }
  Y_l ( \xi_1 , \dots \xi_p ) =
 - Q_l   Y_l ( \xi_1 , \dots , \xi_p ),
 \ee
 where $Q_l$ is the eigenvalue of the Laplace operator on  unit
 $p$ sphere. Possible values of $Q_l$ are $l(l+p-1)$ with $l=0,1,2,
 \cdots$ . Choosing the harmonic oscillation,  perturbations can be expressed  as

 \bea
 \delta \rho ( \tau,  \xi_1 , \dots , \xi_p )
 &=& \tilde \delta \rho e^{- i \omega \tau}
 Y_l ( \xi_1 , \dots , \xi_p ) , \label{delr} \\
 \delta \phi ( \tau,  \xi_1 , \dots , \xi_p )
 &=& \tilde \delta \phi e^{- i \omega \tau}
 Y_l ( \xi_1 , \dots , \xi_p ) ,  \label{delp} \\
 \delta x_k ( \tau,  \xi_1 , \dots , \xi_p )
 &=& \tilde \delta x_k e^{- i \omega \tau}
 Y_l ( \xi_1 , \dots , \xi_p ) .   \label{delx}
 \eea
From Eq. (\ref{eofmxk}), we find the natural frequency for $\delta
x_k$ perturbations as
 \be
 \omega_x^2 = { Q_l \over {(f_0 r_0)^2} } =
 Q_l \omega_0^2 \to \omega_x = \pm \sqrt {Q_\ell}\omega_0.
 \ee
And $\delta \rho$ and $\delta \phi$ perturbations are coupled and
their normal frequencies are determined by the matrix equation

 \be
 \pmatrix{{ 1 \over {(1 - \rho_0^2)}}
 ( - {\omega^2 \over \omega_0^2} + Q_l)&
 i \omega (p-1) {1 \over \omega_0 \rho_0} \cr
  - i \omega (p-1) {1 \over \omega_0 \rho_0}
 & { {(1 - \rho_0^2)} \over \rho_0^2}
 ( - {\omega^2 \over \omega_0^2} + \rho_0^2 Q_l) \cr}
 \pmatrix{ \tilde \delta \rho \cr \tilde \delta \phi \cr}
 ~=~0.
 \ee
This gives the frequencies of two modes ($\pm$)

 \be
 \omega_{\pm} = { 1 \over { 2 (f_0 r_0)^2} }
 \bigg [ ( 1 + \rho_0^2 ) Q_l + (p - 1)^2
 \pm \sqrt{ (p-1)^4 + 2 (p-1)^2 (1 + \rho_0^2) Q_l
       + Q_l^2 (1 - \rho_0^2)^2 } \bigg ].
 \ee

These modes oscillate with real and positive $\omega_{\pm}$, so
their vibration is stable for any size of the probe brane. To make
a connection with ${\rm AdS}_m \times {\rm S}^n$, we can choose
the possible maximum value of $\rho_0$ as $\rho_0=1$. Then the
natural frequencies  for this case are
 \be
 \omega_{\pm} = { 1 \over {  (f_0 r_0)^2} }
 \bigg [ Q_l + { (p - 1)^2 \over 2}
 \pm (p-1) \sqrt{ Q_l + {(p-1)^2 \over 4} } \bigg ].
 \ee
This agrees with the result of the case whose background geometry
is of the form ${\rm AdS}_m \times {\rm S}^n$ (see, Eq.(4.17) of
\cite{djm0009019} and Eq.(73) of \cite{0010206}). Inserting $Q_l =
l(l+p-1)$, we obtain  $\omega_+ = (1/f_0 r_0)(l+p-1)$ and
$\omega_- =  (1/f_0 r_0) l$. Since $l$ is integer, their  motion
is periodic. Here we would like to emphasize that Eq. (\ref{tvn})
is the crucial condition in our calculation. It has been discussed
in \cite{dtv0008203} that we can draw the giant graviton picture
even in non-supersymmetry preserving backgrounds whenever this
condition is met.

It is worthwhile commenting that the case considered in this paper
is somewhat different from that in ${\rm AdS}_m \times {\rm S}^n$
space. Extended brane configurations in ${\rm AdS}_m \times {\rm
S}^n$ space, relevant for the giant graviton picture, sit at
definite value of $r$ both for branes extended in ${\rm AdS}$
subspace and for branes extended on the sphere in which case the
brane sits at the origin of the ${\rm AdS}$ space.
 In our calculation we did not
consider the brane motion along the transverse direction ($r$) in
the background D$(6-p)$-branes. Interestingly it is known that the
brane motion in this direction induces a cosmological evolution on
the universe brane, called the mirage cosmology
\cite{migcos,youm}. Recently it has been studied that the motion
of giant graviton is related to the closed universe of mirage
cosmology\cite{youm}. Hence the motion of a probe $p$-brane along
the $r$-direction is expected to induce an interesting
cosmological evolution in the background whose geometry is not
${\rm AdS}_m \times {\rm S}^n$ space. If we remove the restriction
$\dot r =0$, then the only change might be the existence of the
additional vibrational modes along the $r$ direction which do not
mix with the modes considered here. It would be interesting to
study the vibration modes without the restriction $\dot r=0$.

In summary, we found the normal modes of small
vibrations of giant gravitons in the background of dilatonic
D-branes whose geometry is not of the form ${\rm AdS}_m \times
{\rm S}^n$ and these vibrations are stable.

\section*{Acknowledgement}
Research by J.Y. Kim was  supported in part by the Korea Science
and Engineering Foundation, No. 2001-1-11200-003-2. Research by
Y.S. Myung was supported in part by the Brain Korea 21 Program,
Project No. D-0025.

\end{document}